\documentclass[iop]{emulateapj}

\usepackage{natbib}
\usepackage{amsmath, amssymb}
\usepackage{times}
\usepackage{apjfonts, graphicx, graphics}
\usepackage[breaklinks,colorlinks,urlcolor=blue,citecolor=blue]{hyperref}
\usepackage[all]{hypcap}
\usepackage{aas_macros}
\usepackage{subfigure}
\usepackage{color}
\usepackage{relsize}

\usepackage{subfigure}

\definecolor{grey}{rgb}{0.4,0.5,0.6}
\shorttitle{Spectroscopic tracer vs optical tracers}
\shortauthors{}

\begin{document}
	\title{On the reliability of photometric and spectroscopic tracers of halo relaxation
	}
	\author{Mohammad H. Zhoolideh Haghighi$^{1}$}
	\author{Mojtaba Raouf$^{2,1}$}
	\author{Habib. G. Khosroshahi$^{1}$}
	\author{Amin Farhang $^{1}$}
	\author{Ghassem Gozaliasl $^{3}$}
	
	\affil{
		$^{1}$School of Astronomy, Institute for Research in Fundamental Sciences (IPM), Tehran, 19395-5746, Iran \\
		$^{2}$Korea Astronomy and Space Science Institute, 776 Daedeokdae-ro, Yuseong-gu, Daejeon 305-348, Korea\\
		$^{3}$Department of Physics, University of Helsinki, P. O. Box 64, FI-00014 , Helsinki, Finland\\}

	\begin{abstract}
		We characterize the relaxation state of galaxy systems by providing an assessment of the reliability of the photometric and spectroscopic probe via the semi-analytic galaxy evolution model.  
		We quantify the correlations between the dynamical age of simulated galaxy groups  and popular proxies of halo relaxation in observation, which are mainly either spectroscopic or photometric. We find the photometric indicators demonstrate a stronger correlation with the dynamical relaxation of galaxy groups compared to the spectroscopic probes.
		We take advantage of the Anderson Darling statistic ($A^2$) and the velocity-segregation ($\bar{\Delta V}$) as our spectroscopic indicators, and use the luminosity gap ($\Delta m_{12}$) and the luminosity de-centring ($D_{\mathrm{off-set}}$) as photometric ones. Firstly, we find that a combination of $\Delta m_{12}$  and $D_{\mathrm{off-set}}$ evaluated by a bivariant relation ($\mathrm{B} = 0.04 \times \Delta m_{12} - 0.11 \times Log(D_{off-set}) + 0.28$), shows a good correlation with the dynamical age compared to all other indicators. 
		Secondly, by using the observational X-ray surface brightness map, we show that the bivariant relation brings about some acceptable correlations with X-ray proxies. These correlations are as well as the correlations between $A^2$ and X-ray proxies, offering reliable yet fast and economical method of quantifying the relaxation of galaxy systems. 
		This study demonstrates that using photometric data to determine the relaxation status of a group will lead to some promising results that are comparable with the more expensive spectroscopic counterpart.
	\end{abstract}

	\altaffiltext{*}{
		\href{mailto:mzhoolideh@ipm.ir}{mzhoolideh@ipm.ir}
	}
	
	\section{Introduction}
	
	Observations have revealed that galaxy environments such as groups and clusters influence the properties of the member galaxies \citep{Hashimoto1998, Hou2013}. Nearly half of the galaxy population inhabit in groups \citep{Eke}. They depute an intermediate-mass regime in which a significant population of both spirals and ellipticals are observed \citep{McGee2011}. In the hierarchical structure formation paradigm, groups are progenitors of galaxy clusters; thus their role in galaxies evolution is more significant than it appears \citep{Peng2015, Ellison2008}. For example, galaxy interaction and merger in a group environment can be more efficient because of a lower galaxy velocity compared to the same in a cluster environment, which would influence the morphology of galaxies \citep{Barnes1989}, as well as, their star formation \citep{Wilman2005}. It has been shown that galaxies in dynamically relaxed groups tend to be redder than galaxies in unrelaxed groups \citep{Carollo2013, Raouf2019b} and can grow, through mergers, to become giant galaxies which are mostly found in massive galaxy clusters \citep{Khosroshahi2006b}.
	
	Dynamically relaxed groups and clusters display symmetric X-ray surface brightness distribution \citep{Parekh2015A}, and they often have the brightest central galaxy which is located close to the peak of the X-ray emission e.g. the bottom of potential well \citep{Khosroshahi2004, Skibba2011}. The velocity distribution of member galaxies in a relaxed group is also expected to be Gaussian \citep{Hou2009}. On the other hand, unrelaxed or evolving galaxy groups and clusters are being formed and have been undergoing mergers. 
	
	Recent studies have highlighted the importance of the dynamical state of group size halos in the observed properties of galaxies. For instance, AGN activities of luminous elliptical galaxies probed by their radio emission are shown to be correlated with the dynamical state of the parent halo \citep{Khosroshahi2017}. The brightest group galaxies(BGGs) in the late formed groups are an order of magnitude more luminous in radio than their counterparts in early formed groups which is also confirmed by numerical simulations\citep{Raouf2018}. This follows several observational and numerical studies suggesting that the impact of the group environment on galaxies goes beyond simple overdensity indicators. Therefore, the dynamical state of a galaxy group and its relaxation state plays some role in galaxy evolution.
	
	Fossil galaxy groups are arguably the extreme case of the dynamical relaxation in which luminous galaxies in groups merge to form a giant elliptical galaxy with the halo X-ray emission remaining as the fossil record of such an internal evolution \citep{Ponman1994, Jones2003, Khosroshahi2006a, Khosroshahi2007}. The N-body cosmological simulations support this argument \citep{D'onghia2005, Dariush2007, Raouf2014}.
	With the advances in the imaging surveys, many current and future large imaging surveys will benefit from reliable and economical methods of assigning \textit{the age dating galaxy groups} introduced by \citep{Raouf2014}. In the following studies, \citet{Lavoie2016}, \cite{Horellou2018}, and \cite{Gozaliasl2019} used the luminosity gap between two most luminous galaxies in the group, $\Delta m_{12}$, and/or the off-set between brightest group galaxies and the luminosity centroid, $D_{\mathrm{off-set}}$ to probe relaxation and dynamical age of halos.  
	
	There are several approaches for parametrizing halo relaxation. If the X-ray data be available, the photon asymmetry \citep{Nurgaliev2013} or centroid shift \citep{Bohringer2010} can be used as indicators. In the case of photometric data, the indicators are generally the luminosity gap \citep{Jones2003} and the most BGG offset \citep{Sanderson2009}. Also, some spectroscopic indicators could be used; such as, velocity-segregation \citep{Raouf2019b, Lares, Nascimento} and Anderson Darling (AD) test which is based on the Gaussianity of the velocity distribution \citep{Hou2009} of group galaxies. It is worth mentioning that since X-ray observations are both more expensive and less available than spectroscopic or photometric observations, some suggest that it would be better to use AD test instead of X-ray probes \citep{Roberts2018}. They come to this conclusion by investigating the correlation between X-ray indicators and AD test and showing that these quantities correlate strongly.
	
	In this study, we investigate the reliability of photometric and spectroscopic and the X-ray proxies to determine the dynamical state of galaxy groups. In the era of large surveys,  reliable and yet economical method of characterizing the halo dynamical state would be of great advantage for statistical studies aimed at understanding \textit{the role of environment on galaxy properties}. Motivated by our previous studies employing the luminosity gap and the BGG offset, we introduce a bivariant correlation which is a combination of $\Delta m_{12}$ and $D_{\mathrm{off-set}}$ to test whether we can overcome the superiority of the AD test, $A^2$, which is explored by \citet{Hou2009, Hou2013}. In our study, we use galaxies drawn from the semi-analytic models of {\cite{Raouf2017}}, based
	on the Millennium Simulation \citep{Springel2005}, so-called Radio-SAGE. The advantage of simulated data is that through the semi-analytic models we can reliably obtain the mass assembly history of dark matter halo. We also apply our finding on observed galaxy groups of the Yang catalog \citep{Yang2005,Yang2007} to compare the predictions of the simulations with the observations. Throughout this paper, we adopt  $H_0 = 100 h\,\rm km\,s^{-1}\, Mpc^{-1}$ for the Hubble constant with $h=0.73$.

	\section{Data and mock catalog} \label{Analysis}
	\subsection{ Simulated data}
	
	In this study we use the public release of the Millennium Simulation with a $\Lambda$CDM  cosmology and the following parameters: ${\Omega}_m=0.25$, ${\Omega}_b=0.045$, ${\Omega}_{\Lambda}=0.75$, h=0.73, n=1,${\sigma}_8=0.9$. The simulation box $(500 h^{-1} Mpc)^{3}$ contains $2160^{3}$ particles and presents the mass resolution of 8.6${\times}$ 10$^8$ ~h$^{-1}$~M$_{\odot}$.  The dark matter merger trees within each simulation snapshot (64 snapshots) are expanded approximately logarithmically in time between $z=127$ and $z=0$ and extracted from the simulation using a combination of friends-of-friends (FoF) \citep{Davis1985} and SUBFIND \citep{Springel2001} halo finders algorithm. The gas and stellar components of galaxies in dark matter halos are constructed semi-analytically, based on different phenomenological recipes. We are using the Radio semi-analytic galaxy evolution \citep[][Radio-SAGE]{Raouf2017} galaxy formation model that self consistently follows the gas cooling-heating cycle in a different type of galaxies and calibrated to match key observations for various redshift \citep{Raouf2019a}.  The galaxy catalog contains  $\sim 51000$ halos with masses above 10$^{13}$ ~h$^{-1}$~M$_{\odot}$  and    $\sim5$ million galaxies from which we only select galaxies brighter than $-14$ in r-band absolute magnitude for completeness. Also, we have chosen systems with $Mr_{BGG} < -21.5$ to remove the modest galaxies with dwarf satellites.
	
	\subsection{Mock redshift-space catalog} \label{mock}
	
	To take into account the basic observational limitations, a mock catalog has been constructed from the algorithm described in \cite{Blaizot} without box transformations or replication. To do so we (i) place the observer at one of the vertices of the simulation box and look at the galaxies in the box through the observer line of sights, (ii) then convert the Cartesian coordinate system (i.e. X, Y, and Z) to celestial coordinates (i.e., R.A, and DEC.), (iii) and measure the redshift of each galaxy using \cite{Duarte} algorithm, (iv) estimate the luminosity distance D$_{L}$ of each galaxy, using their computed redshift and comoving distance, (v) the apparent magnitude of each galaxy is then computed from D$_{L}$ and the absolute magnitude. We estimate the uncertainties on the mean of our measurements in the same way that is described by \cite{Farhang}, sec~4.1. Finally, having the redshift of the objects, we can easily calculate the line of sight velocity of group members.
	
	\subsection{Observational data}
	
	To have a fair comparison with the findings of \cite{Roberts2018}, we select the same sample and the same data that they adopt. Therefore we use galaxy groups/clusters with halo mass $M_{halo} > 10^{13}$ provided by the Yang catalog \citep{Yang2005,Yang2007} which are recognized through the friend-of-friend algorithm \citep[FoF;][]{Huchra1982,press1982}.
	To compute the cluster-centric radii we use galaxies redshift, the angular separation between the galaxy position and the luminosity-weighted center of the cluster and then we normalize it to $R_{500}$  which is the radii that the inner average density is 500 times the critical density  of the Universe.

	 A subset of the Yang catalog comprising clusters with a minimum of ten spec-z  members is chosen to ensure an accurate classification, using the velocity profile shape \citep{Hou2009}.  To study the relation between optical and X-ray relaxation indicators, We use a sample of 58 clusters by \cite{Roberts2018}, which are found after the cross-matching Yang catalog with the Chandra and \textit{XMM-Newton} X-ray observations archives. Only X-ray observations with clean exposure times $\geq$ 10 ks have been chosen. For the Chandra images, observations were reprocessed, cleaned, and calibrated by {\scriptsize CIAO} version 4.9, and {\scriptsize CALDB} version 4.7.5. Also {\scriptsize LC$\_$CLEAN } with a $3\sigma$ threshold is used to filter background flares. Besides, charge transfer inefficiency and time-dependent gain corrections are taken into account. Images are created in the 0.5$-$5 keV energy band. By using the {\scriptsize WAVDETECT} script, point sources are identified and then filled with local Poisson noise using {\scriptsize DMFILTH}. Blank sky background images are generated for each observation using the {\scriptsize BLANKSKY} and {\scriptsize BLANKSKY$\_$IMAGE} scripts.  For \textit{XMM–Newton}, data reduction observations are done by {\scriptsize SAS}, version 16.0.0. 
	Calibrated event files are generated using the {\scriptsize EMCHAIN} script, and filtered event lists were generated using {\scriptsize MOS-FILTER}. Like the {\scriptsize CIAO} images, exposure corrected images are created in the 0.5$-$5 keV band, and point sources are identified with the {\scriptsize CHEESE} script and thereafter filled with local Poisson noise using the {\scriptsize CIAO} script {\scriptsize DMFILTH}. For more details on the X-ray data reduction, we refer the reader to \cite{Roberts2018}.
	
	\section{Mass assembly history and Halo relaxation proxies}
	
	\subsection{Mass assembly history}
	
	We assign a dynamical age to each cluster by obtaining the halo mass ratio at $z \sim 0.5,1$ to $z \sim 0$ which we specify it with $\alpha_{z,0} = M_{200}(z = 0.5,1)/M_{200}(z = 0)$.  We define fast growth and slow growth modes associated with $\alpha_{0.5,0}$ and $\alpha_{1,0} $ respectively. Our intution is based on the fact that there would be less available time for halos to grow from $z=0.5$ compared to those halos growing from $z=1$. According to our definition \citep{Dariush2007, Raouf2014}, a group is dynamically unrelaxed if it reaches less than one-third of its final mass by $z \sim 1$ ($\alpha_{1,0}< 0.3$) and is relaxed if it reaches  more than half of its total present-day mass by $z \sim 1$ ($\alpha_{1,0}> 0.5$). In our earlier studies \citep{Raouf2016}, we showed a sample of relaxed groups selected based on the luminosity gap and the BGG/BCG offset from the halo center 
	will result in a contaminated sample with high dynamical age ($\alpha_{1,0}>0.5$). Nevertheless, we note that this method for estimating the dynamical age is not unique and may show different result when we track the halo individually \citep[e.g see trace method and scatters in; ][]{Dariush2010,Gozaliasl2014,Farhang}.
	
		\subsection{Tracers based on the luminosity distribution}
		\subsubsection{luminosity gap}
	\label{Lgap}
	
	One of the most promising and successful halo age indicators is the luminosity gap ($\Delta m_{12}$). It is the magnitude difference between the first and second brightest galaxy within half the virial radius of a group as introduced by \cite{Jones2003} when it's larger than 2 mag for conventional definition the Fossil galaxy groups \citep{Ponman1994}. 
	Alternatively, some authors prefer to use the stellar mass ratio between the second most-massive and most-massive galaxies in a given group \citep{Roberts2018}. 
	For a relaxed system, the luminosity gap is large then the stellar mass ratio $M_2/M_1$ should be small, while for younger groups $M_2/M_1$ hasn't reduced enough and is not so small.
	
	\subsubsection{Luminosity de-centring} 
	The BGG is expected to be located at the center of the group's halos if the group is dynamically relaxed  \citep{Ponman1994, smith2005}.  
	We are using optical luminosity de-centering, $D_{\mathrm{off-set}}$ as a tracer of the dynamical age of the galaxy groups. Merging systems are unrelaxed and have their BGG is displaced from the center of the group halo. To find the position of the halo center alternatively, one can use the X-ray peak and the mass centroid from the gravitational lensing observations \citep{Oguri2010,Dietrich2012, Gozaliasl2019}.  Given that the lensing mass map and the X-ray emission peaks are not directly accessible through cosmological simulations, we rely on the luminosity weighted centroid of galaxy groups which is also economically available in the optical observations.
	
	We calculate luminosity weighted/centroid by using: $ X_L = \Sigma X_i L_i /L_i$, where $L_i$ is the r-band luminosity of the $i$th galaxy in a group in and $X_i$ is the projected coordinate of each galaxy.

	\subsubsection{Bivariant correlation ($\mathrm{B}$)}\label{sec-bivariant}
	
	In order to come up with a more appropriate photometric probe, we combine $\Delta m_{12}$ and $D_{\mathrm{off-set}}$ into a linear bivariant correlation as: 
	\begin{equation}\label{eq:B}
	   	\mathrm{B} = C \times\Delta m_{12} + D \times Log(D_{off-set}) + E 
	\end{equation}
	This bivariant correlation is basically a mixture of both centroid shift and luminosity gap. By studying the relation of $\mathrm{B}$ with $\alpha_{z,0}$ and minimizing the scatter between these two quantities, we can determine the constant coefficients ($C, D, E$) of the above relation. Specifically, these constants can be found via the least square method, after which we calculate the correlation coefficient between $\alpha_{z,0}$ and $\mathrm{B}$.
	%\section{Spectroscopic probes  of Halo relaxation}
	
		\subsection{Dynamical tracers}
		\subsubsection{Anderson Darling (AD) test}
	
	Some studies \citep{Yahil1977,Ribeiro2013} show that the line-of-sight velocity distribution of member galaxies within a relaxed group/cluster is almost normal; however, the unrelaxed groups display a larger deviation from normal velocity distribution.
	We can measure the deviation of velocity distribution from normality by using Kolmogorov, $\chi^2$, or Anderson Darling (AD) tests. Since the AD test has been shown to be more powerful and reliable than other tests in detecting departures from an underlying Gaussian distribution \citep[see;][for details and uncertainties]{Hou2009}, we employ this test to measure \textit{how} much a velocity distribution of member galaxies deviates from a normal distribution.
	
	The AD test relays on calculating the distance between the cumulative distribution functions (CDFs) of an specific distribution and an ideal normal distribution. This distance can be measured in terms of $A^2$ according to the following relation:

	\begin{equation}
	A^2 = -n - \frac{1}{n} \Sigma_{i=1 }^n [2i-1][\ln \Phi(x_i) + \ln(1-\Phi(x_{n+1-i}))],
	\label{A2}
	\end{equation} 
	where $x_i < x < x{_i+1} $ and $\Phi(x_i)$ is the CDF of the hypothetical underlying distribution. Large values of this statistic ($A^2$) correspond to larger deviations from normality.
	In the case of a Gaussian distribution which is what we are considering here we have:
	
	\begin{equation}
	\Phi(x_i) = \frac{1}{2}\Big(1+ erf(\frac{x_i - \mu}{\sqrt{2} \sigma})\Big).
	\label{phi}
	\end{equation}
	By calculating $A^2$ for a distribution with an arbitrary significance and comparing it with critical values we can conclude if a distribution is normal or not \citep{Stephens1974}.

		\subsubsection{Velocity-segregation ($\bar{\Delta V}$)}
	
	Another spectroscopic indicator which is helpful to determine the relaxation state of the galaxy groups is the velocity-segregation. We calculate the velocity segregation  between the BGG and the $i$th spectroscopic member galaxy within half virial radius, using the following relation:  
	\begin{equation}
	\bar{\Delta V} = \frac{\Sigma_{i=1}^{n-1}|V_{BGG}-V_i|}{n-1},
	\end{equation}
	where $n$ is the number of galaxies in a group, $V_{BGG}$ and $V_{i}$ are the line of sight velocities of the Brightest Group Galaxy and $i$th galaxy respectively.  Motivated from the velocity profile of groups and clusters, $\bar{\Delta V}$ should be smaller for relaxed systems; in contrast, it should be large for dynamically unrelaxed ones \cite[see fig. 11; ][]{Raouf2019b}.

	%\section{The X-ray probes  of Halo relaxation}\label{sec:x-ray:indic}
		\subsection{ICM tracers} \label{sec:x-ray:indic} 
		\subsubsection{Photon Asymmetry ($A_{phot}$)}
	
	Photon asymmetry is one of the best model-independent and most robust techniques to measure the asymmetry of the X-ray profiles \citep{Nurgaliev2013}. This novel method quantifies the degree of axisymmetry of X-ray photon distributions around the X-ray peak, in other words, it demonstrates how uniform photons are placed in a $2 \pi$ radian range within an annulus encompassing the cluster center. For a detailed explanation regarding  the photon asymmetry we refer the reader to  \citep{Nurgaliev2013}. We take advantage of the photon asymmetry calculated by \cite{Roberts2018}  using the following equations \ref{eq:aphot}:
	
	\begin{equation}
	A_{phot} = 100 \frac{\Sigma_{j=1 }^{N_{ann}} C_j \hat{d}_{N_{j},C_{j}}}{\Sigma_{j=1 }^{N_{ann}} C_j },
	\label{eq:aphot}
	\end{equation}
	where $N_{ann}$ is the total number of annuli, $C_{j}$ is number of cluster counts above background within $j$-th annulus and $\hat{d}_{N_{j},C_{j}}$ is an estimated distance between the true photon distribution and the uniform distribution, given as follow:
	\begin{equation}
	\hat{d}_{N,C} = \frac{N}{C^2}\big(U^2_N - \frac{1}{12}\big),
	\end{equation}
	where $N$ is the total number of counts within each annulus, and $C$ is the number of counts intrinsic to the cluster. Besides, $U^2_N$ is Watson's statistic \citep{WATSON}, which can be achieved by minimizing the following equation:
	
	\begin{equation}
	U^2_N(\phi_0) = \frac{1}{12N} + \Sigma_{i=0 }^{N-1}\big(\frac{2i+1}{2N} - F(\phi_i) \big)^2 - N \big(\frac{1}{2} - \frac{1}{N}\Sigma_{i=0 }^{N-1} F(\phi_i) \big)^2,
	\end{equation}
	here  $\phi_i$  are the observed count polar angles, $\phi_0$ is the origin polar angle on the circle, and $F$ is the uniform CDF.
	
	\begin{equation}
	U^2_N  =  \min_{origin ~ on~circle, \phi_0} U^2_N(\phi_0) 
	\end{equation}
	Adapting a similar approach as discussed in \cite{Nurgaliev2013}, $ \hat{d}_{N,C} $ is calculated in four radial annuli, which in this study they range between $0.05R_{500}$ and $0.5R_{500}$

		\subsubsection{The centroid shift ($w$)}
	
	Another popular X-ray relaxation indicator is the centroid shift, $w$. It measures the shift of the X-ray surface brightness centroid in different radial apertures. While the ICM center of mass of a system in dynamical equilibrium should scale independently, an un-relaxed system has a scale-dependent center of mass \citep{Mohr}. Therefore, by taking advantage of $w$, relaxation status of a cluster could be determined. Centroid shift can be calculated based on the following relation \citep{Bohringer2010}:
	
	\begin{equation}
	w = [\frac{1}{N-1}\mathlarger{\mathlarger{\Sigma_{i}}} (\Delta_i ~-<\Delta>)^2]^{1/2} \times \frac{1}{R_{max}}.
	\end{equation}
	In the above equation, $\Delta_i$ is the offset between the X-ray peak and the centroid position within the $i$th aperture, N is the number of apertures, and $R_{max}$ is the radius of the largest aperture. X-ray peak is selected to be the position of the brightest pixel, and centroids are specified from the moments of the exposure-corrected X-ray images. Nine apertures are chosen in the range of $0.1 R_{500}$ to  $0.5 R_{500}$ with a   $0.05 R_{500}$ step size.

	\section{Results}
	\subsection{Relation between $\alpha_{z,0}$ and different relaxation proxies}
	To compare different relaxation probes, introduced in section 3, and investigate their pros and cons, we follow two steps. Firstly, we fit a power law to the pairs of introduced proxies and $\alpha_{z,0}$ (for z = 1, 0.5), which provides us fitting parameters. Secondly, by evaluating Spearman's rank correlation coefficient, we quantify the degree of correlation between each pair. In the following subsections, we study the correlation between different relaxation probes and $\alpha_{z,0}$ in more detail.
	
	\subsubsection{Correlation between $\alpha_{z,0}$ and photometric probes }
\begin{figure}
		\centering     %%% not \center
		\subfigure{\includegraphics[width=42mm]{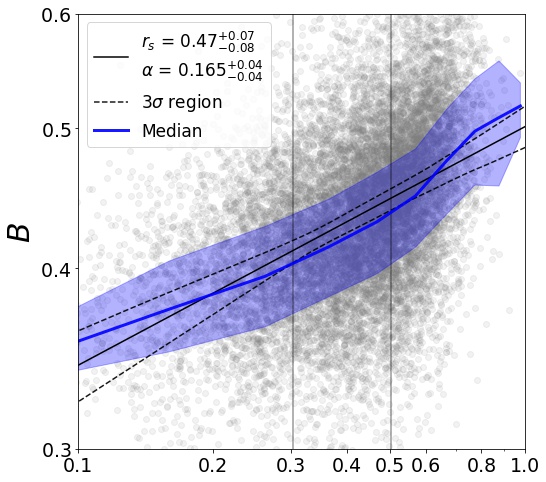}}
		\subfigure{\includegraphics[width=42mm]{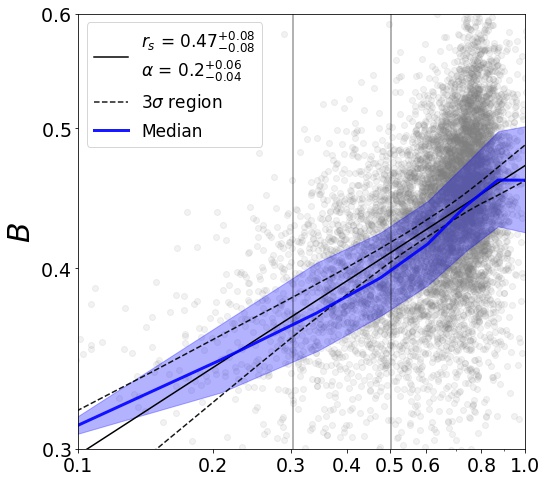}}
		\subfigure{\includegraphics[width=42mm]{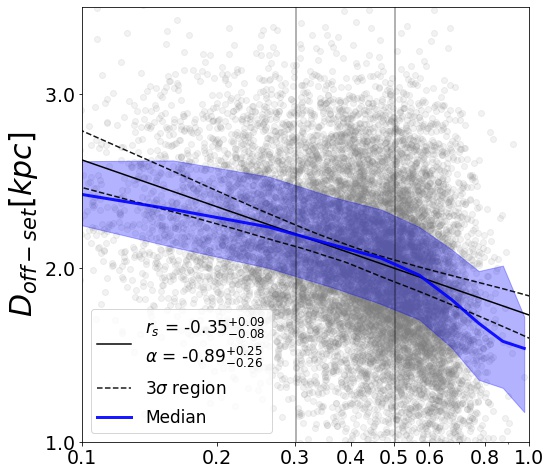}}
		\subfigure{\includegraphics[width=42mm]{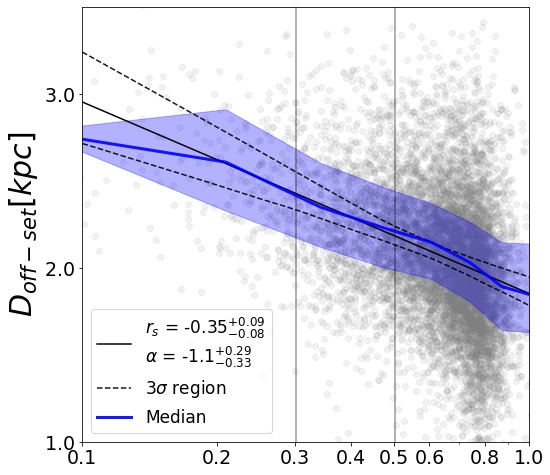}}
		\subfigure{\includegraphics[width=42mm]{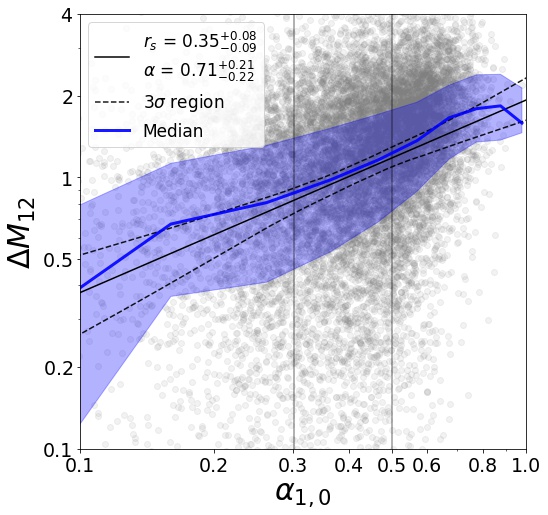}}
		\subfigure{\includegraphics[width=42mm]{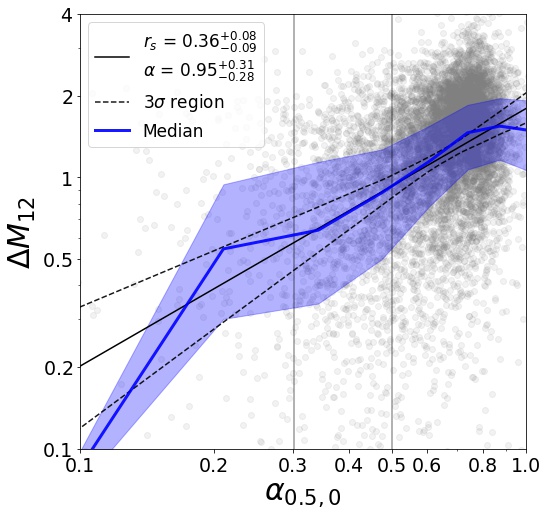}}
		\caption{ Correlation between photometric indicators ($\Delta M_{12}$, $D_{\mathrm{off-set}}$, $\mathrm{B}$)  and the halo mass ratio at different redshifts ($\alpha_{1,0}$, $\alpha_{0.5,0}$). The Spearman's correlation coefficient $r_s$ and slope is reported in each panel. Two vertical lines indicate the dynamically relaxed ($\alpha_{1,0}>0.5$) and  unrelaxed groups ($\alpha_{1,0}<0.3$) \citep{Raouf2014}. The  area between dashed lines is showing a $3 \sigma$ confidence intervals of our fits. The first and third quartile ($Q_1 \& Q_3$) are illustrated in the shaded regions.} \label{fig:age1}
	\end{figure}

	In Figure \ref{fig:age1}, we show the photometric indicators versus $\alpha_{z,0}$. The x-axis of all left-hand side plots is based on $\alpha_{1,0}$; however, the right panels' x-axis is based on $\alpha_{0.5,0}$. Both the best fit line with $3\sigma$ confidence interval and median of $\Delta M_{12}$, $D_{\mathrm{off-set}}$, and $\mathrm{B}$ are demonstrated in each panel (from bottom to top). We have taken advantage of the bootstrap method with 10000 random sampling with replacement to calculate these confidence intervals. additionally, not only all plots comprise scattered data, but also they involve the first(0.25) and third(0.75) quartiles ($Q_1\  \& Q_3$), shown by the shaded regions.
		
	In the bottom panel, we show the correlations between $\Delta M_{12}$ and $\alpha_{z,0}$, which we have quantified by means of the Spearman's correlation coefficient. 
	Through the relation with $\alpha_{1,0}$, the correlation is positive with $r_s=0.35^{+0.08}_{-0.09}$ and $\alpha=0.71^{+0.21}_{-0.22}$ in which $r_s$ refers to  Spearman's correlation coefficient and $\alpha$ is the slope of the best fit line. Following the similar approach for $\alpha_{0.5,0}$, we end up getting $r_s=0.36^{+0.08}_{-0.09}$.  Comparing the Pearson's coefficient of $\Delta M_{12}-\alpha_{0.5,0}$ and $\Delta M_{12}-\alpha_{1,0}$ relations, we observe no significant difference between the Pearson's correlation coefficient of fast and slow growth; however, their slope and intercept differ slightly.

	A test were run to investigate the range of growth histories that correspond to $\Delta M_{12}$. As can be seen from the bottom left panel, the values of  $\Delta M_{12}$ at $\alpha_{1,0} = 0.3$ and $\alpha_{1,0}=0.5$ are  $\Delta M_{12} = 0.83$ and $\Delta M_{12} = 1.2$ respectively. We consider those halos with $\Delta M_{12} < 0.83$ as unrelaxed and those with $\Delta M_{12} > 1.2$ as relaxed. We find that   $47\%$ of relaxed halos are early-formed ($\alpha_{1,0} > 0.5$) and  $13\% $ are late-formed($\alpha_{1,0} < 0.3$). On the other hand, $17\%$ of unrelaxed halos are early-formed and  $33\% $ are late-formed.
	in view of this, there is a   correlation between $\alpha_{z,0}$ and $\Delta m_{12}$ from which in particular, a large number of of early-formed and late-formed halos could be correctly labeled as relaxed and unrelaxed, respectively.

		Moving on the second photometric proxy,
		 we show distribution of $D_{\mathrm{off-set}}$ versus $\alpha_{z,0}$ in the middle panel of  Figure \ref{fig:age1}. By studying the correlation between $\alpha_{1,0}$ and $D_{\mathrm{off-set}}$, we find  that the two quantities unti-correlated with, $r_s =-0.35^{+0.09}_{-0.08} $ and $\alpha = -0.89^{+0.25}_{-0.26}$. Applying the same approach for $\alpha_{0.5,0}$ bring about $r_s =-0.35^{+0.09}_{-0.08} $. After comparing the Pearson's coefficient of $D_{off-set}-\alpha_{0.5,0}$ and $D_{off-set}-\alpha_{1,0}$ relations, we report no significant difference between the correlation coefficient of fast and slow growth, while their slope and intercept differ slightly.
	
		Similar to the above investigation for luminosity gap, here we investigate the range of growth histories that correspond to $D_{\mathrm{off-set}}$. As can be seen in the middle left part of Figure \ref{fig:age1}, the values of  $D_{\mathrm{off-set}}$ at $\alpha_{1,0}= 0.3$ and $\alpha_{1,0}=0.5$ are  $D_{off-set} = 2.17 kpc$ and $D_{off-set} = 1.97 kpc$ respectively. We consider those halos with $D_{off-set} > 2.17 kpc$ as unrelaxed and those with $D_{off-set} < 1.97 kpc$ as relaxed.   As a result, we see that $48\%$ of relaxed halos have early-formed and  $11\% $ are late-formed. On the other hand, $22\%$ of unrelaxed halos are early-formed and  $31\% $ are late-formed.
		  In summary, there is  an anti-correlation between $\alpha_{z,0}$ and $D_{\mathrm{off-set}}$ from which a large number of early-formed and late-formed halos could be correctly labeled as relaxed and unrelaxed respectively.

	\begin{table}
		\centering
		\begin{tabular}{l l l l l }
			\hline
			Proxy &   slope    & intercept    &  $r_s$        & p-value         \\
			\hline
			$\log(A^2)$    & $-0.30^{+0.21}_{-0.24}$ & $-0.43^{+0.05}_{-0.06}$    & $-0.12^{+0.10}_{-0.09}$  & $0.01$  \\        
			$\log(\bar{\Delta V})$  & $-0.41^{+0.08}_{-0.09}$   &$2.3^{+0.02}_{-0.02}$   & $-0.43^{+0.08}_{-0.07}$   & 0.0  \\
			$\log (D_{off-set})$    & $-0.89^{+0.25}_{-0.26}$      & $1.7^{+0.07}_{-0.07}$      & $-0.35^{+0.09}_{-0.08}$   & 0.0     \\
			$\log(\mathrm{B} )$    & $0.165^{+0.04}_{-0.04}$      & $-0.3^{+0.02}_{-0.01}$      & $0.47^{+0.08}_{-0.07}$    & 0.0    \\    
			$\log(\Delta m_{12})$    & $0.71^{+0.21}_{-0.22}$      & $0.28^{+0.06}_{-0.06}$      & $0.35^{+0.08}_{-0.09}$     & 0.0       \\
			\hline        
		\end{tabular}
		\caption{Correlation between $\log(\alpha_{1,0})$  and different relaxation proxies. $A^2$ and $\bar{\Delta V}$ are spectroscopic indicators, while $D_{\mathrm{off-set}}$, $\mathrm{B}$ and  $\Delta m_{12}$ are photometric indicators. $3\sigma$ confidence intervals are calculated by the bootstrap method.}
		\label{tabel1}
	\end{table}

	\begin{table}
		\centering
		\begin{tabular}{l l l l l }
			\hline
			Proxy &   slope    & intercept    &  $r_s$        & p-value        \\
			\hline
			$\log(A^2)$    & $-0.34^{+0.26}_{-0.31}$ & $-0.43^{+0.05}_{-0.06}$    & $-0.10^{+0.09}_{-0.09}$  & $0.02$  \\        
			$\log(\bar{\Delta V})$  & $-0.50^{+0.10}_{-0.12}$   &$2.3^{+0.02}_{-0.02}$   & $-0.42^{+0.08}_{-0.08}$   & 0.0  \\
			$\log (D_{off-set})$    & $-1.1^{+0.29}_{-0.33}$      & $1.85^{+0.05}_{-0.05}$      & $-0.35^{+0.09}_{-0.08}$   & 0.0     \\
			$\log(\mathrm{B} )$    & $0.20^{+0.04}_{-0.05}$      & $-0.33^{+0.01}_{-0.01}$      & $0.47^{+0.08}_{-0.08}$    & 0.0    \\    
			$\log(\Delta m_{12})$    & $0.95^{+0.31}_{-0.28}$      & $0.25^{+0.04}_{-0.04}$      & $0.36^{+0.08}_{-0.09}$     & 0.0       \\
			\hline        
		\end{tabular}
		\caption{Correlation between $\log(\alpha_{0.5,0})$  and different relaxation proxies. $A^2$ and $\bar{\Delta V}$ are spectroscopic indicators, while $D_{\mathrm{off-set}}$, $\mathrm{B}$ and  $\Delta m_{12}$ are photometric indicators.$3\sigma$ confidence intervals are calculated by the bootstrap method.}
		\label{tabel05}
	\end{table}

	    Finally, in the top panels of Figure \ref{fig:age1}, after calculating the bivariant correlation for the underlying groups, we plot it as a function of $\alpha_{z,0}$. 
	   We recognize that the correlation between $\mathrm{B}$ and $\alpha_{z,0}$ is  significantly stronger than when we use the luminosity gap or de-centering separately.
		The value of the Spearman's correlation coefficient, $r_s$, for $\mathrm{B} - \alpha_{1,0}$ relation is $0.47^{+0.08}_{-0.07}$, and  for $\mathrm{B} -\alpha_{0.5,0}$ turn out to be $0.47^{+0.08}_{-0.08}$.  In addition, the best values for $C, D$ and $E$ found to be $0.04, -0.11$ and $0.28$, respectively.
	
		Same as previous test, here we measure the range of growth histories that correspond to  $\mathrm{B}$.
		As can be seen from top left panel of Figure \ref{fig:age1}, the values of  $\mathrm{B}$ at $\alpha_{1,0}= 0.3$ and $\alpha_{1,0}=0.5$ are  $\mathrm{B} = 0.41$ and $\mathrm{B} = 0.45$ respectively.
		 We consider those halos with $\mathrm{B} < 0.41$ as unrelaxed and those with $\mathrm{B} > 0.45$ as relaxed. Therefore,  we find that $54\%$ of relaxed halos are early-formed and  $10\% $ are late-formed. Also, $16\%$ of unrelaxed halos are early-formed and  $36\% $ are late-formed.
		 As a result, a large number of early-formed and late-formed halos could be correctly labeled as relaxed and unrelaxed respectively.
	
	\begin{figure}
		\centering     %%% not \center
		\subfigure{\includegraphics[width=42mm]{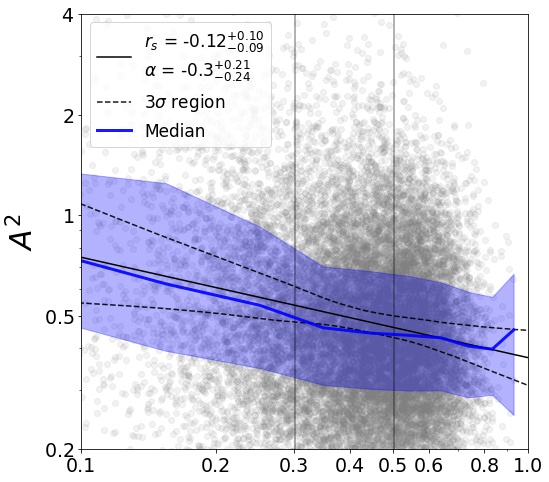}}
		\subfigure{\includegraphics[width=42mm]{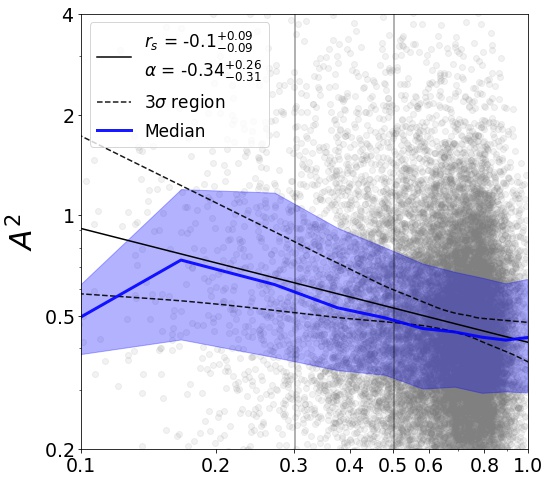}}
		\subfigure{\includegraphics[width=42mm]{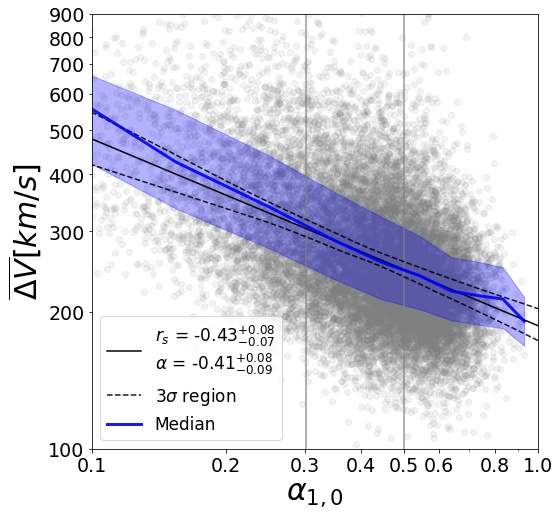}}
		\subfigure{\includegraphics[width=42mm]{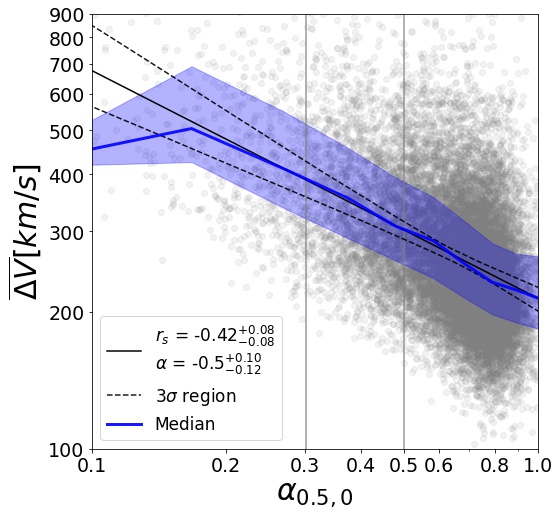}}
		\caption{ Correlation between spectroscopic indicators ($A^2$, $\bar{\Delta V}$) and the halo mass ratio at different redshifts ($\alpha_{1,0}$, $\alpha_{0.5,0}$). The Spearman's correlation coefficient $r_s$ and slope is reported in each panel. Two vertical lines indicate the dynamically relaxed ($\alpha_{1,0}>0.5$) and  unrelaxed groups ($\alpha_{1,0}<0.3$) \citep{Raouf2014}. The  area between dashed lines is showing a $3 \sigma$ confidence intervals of our fits. The first and third quartile ($Q_1 \& Q_3$) are illustrated in the shaded regions.}
		\label{fig:age2}
	\end{figure}

	\begin{figure*}    
		\centering
		\subfigure{\includegraphics[width=4.9cm]{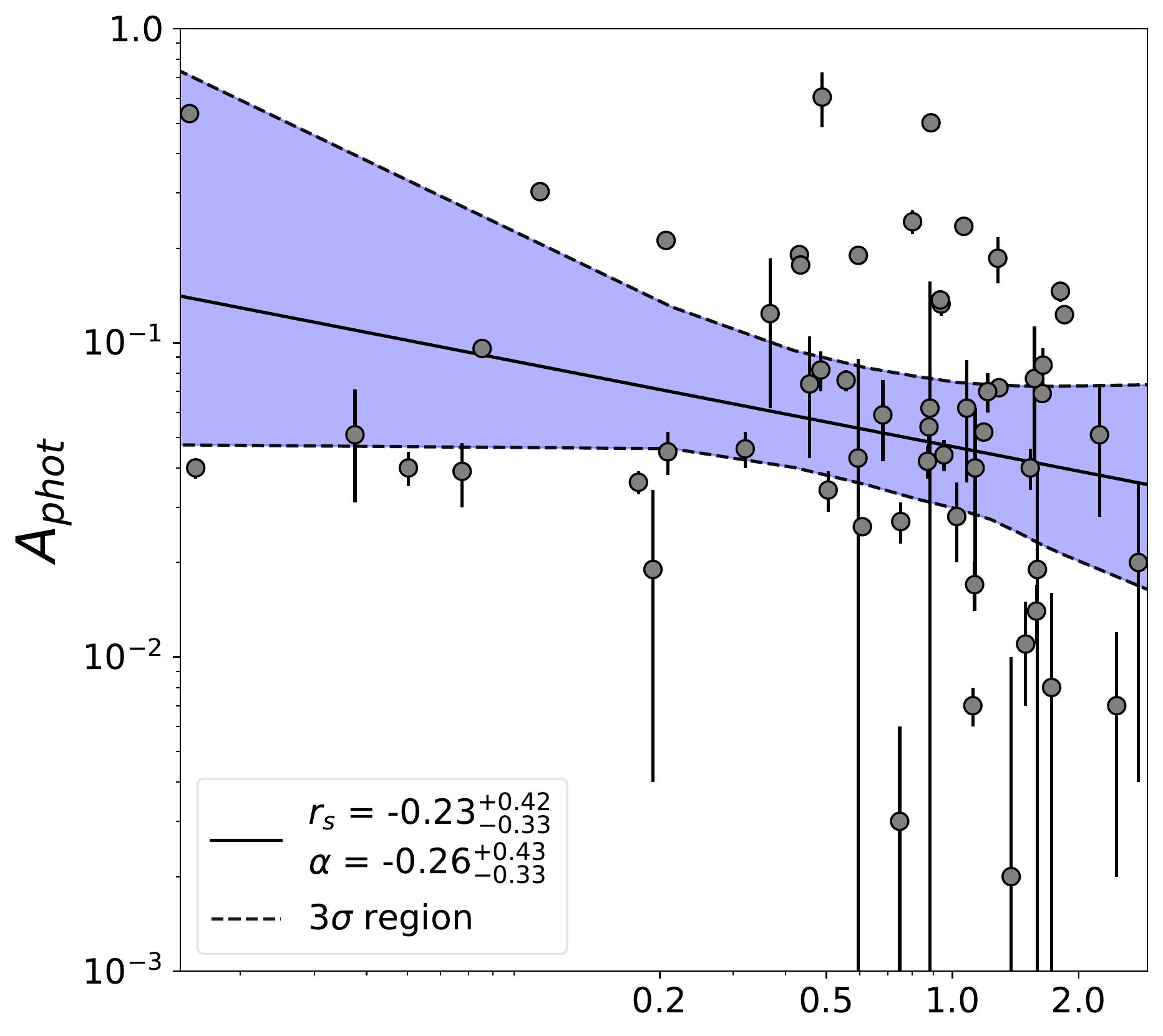}}
		\subfigure{\includegraphics[width=4.9cm]{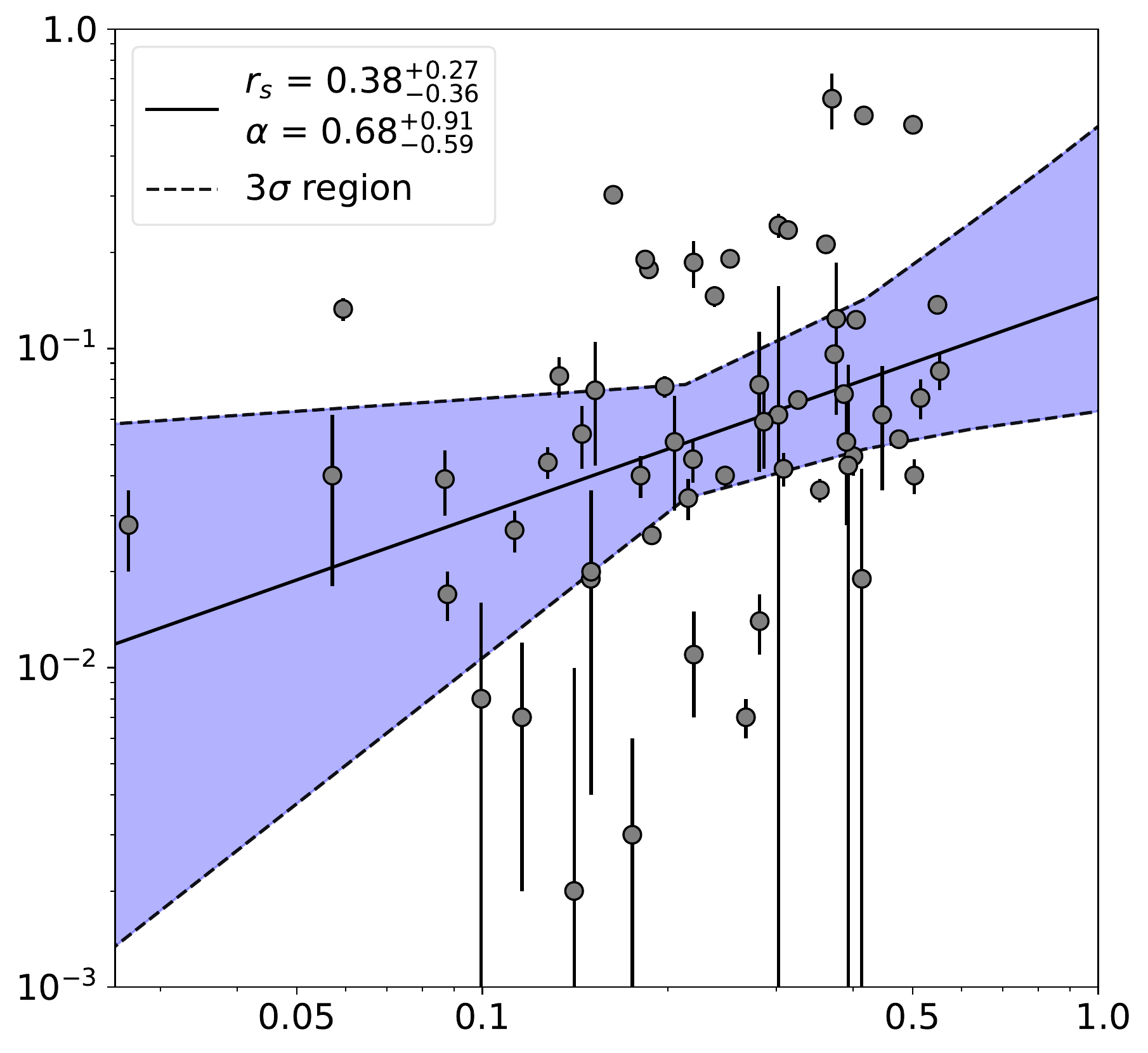}}
		\subfigure{\includegraphics[width=4.9cm]{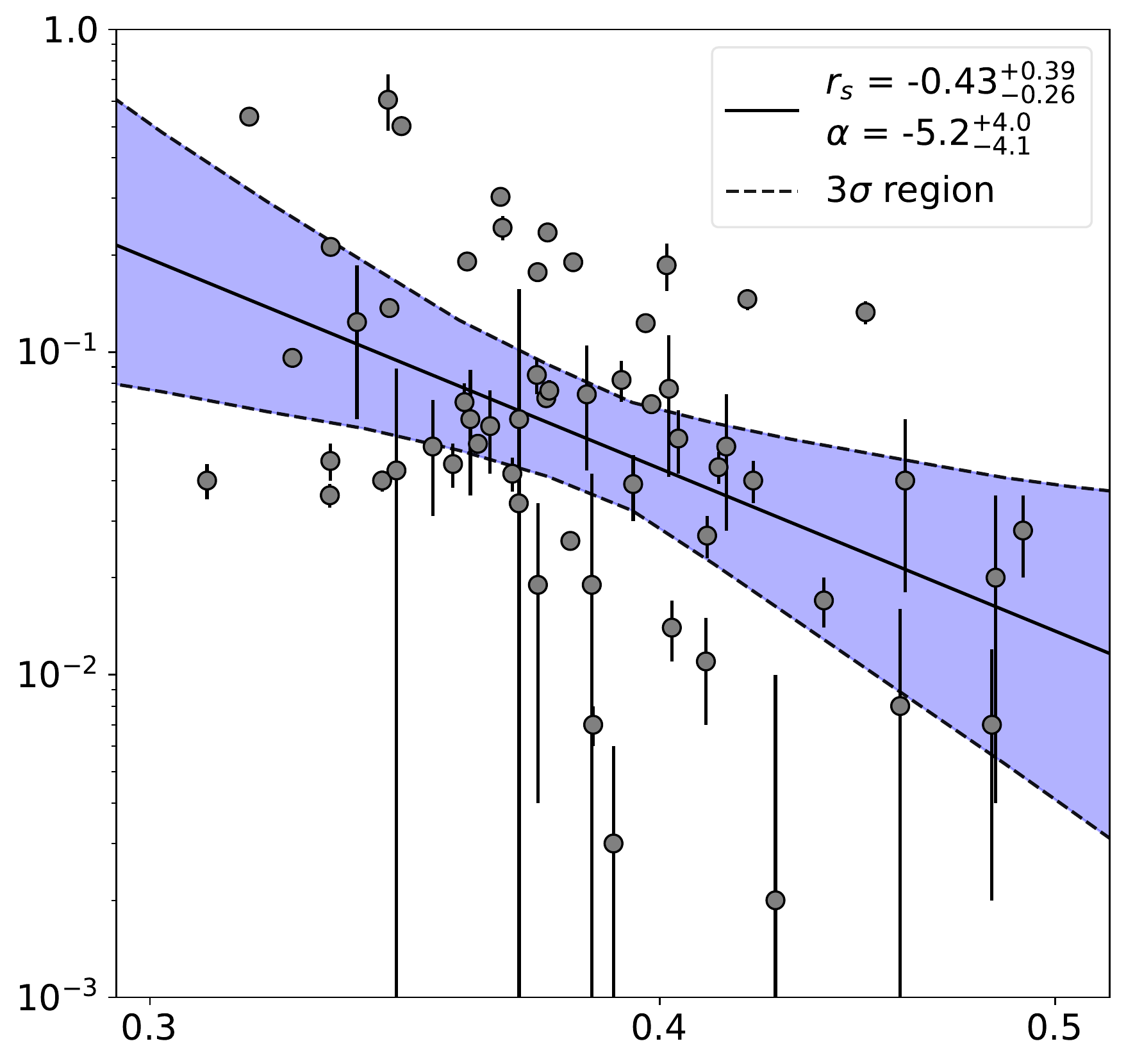}}
		\subfigure{\includegraphics[width=4.9cm]{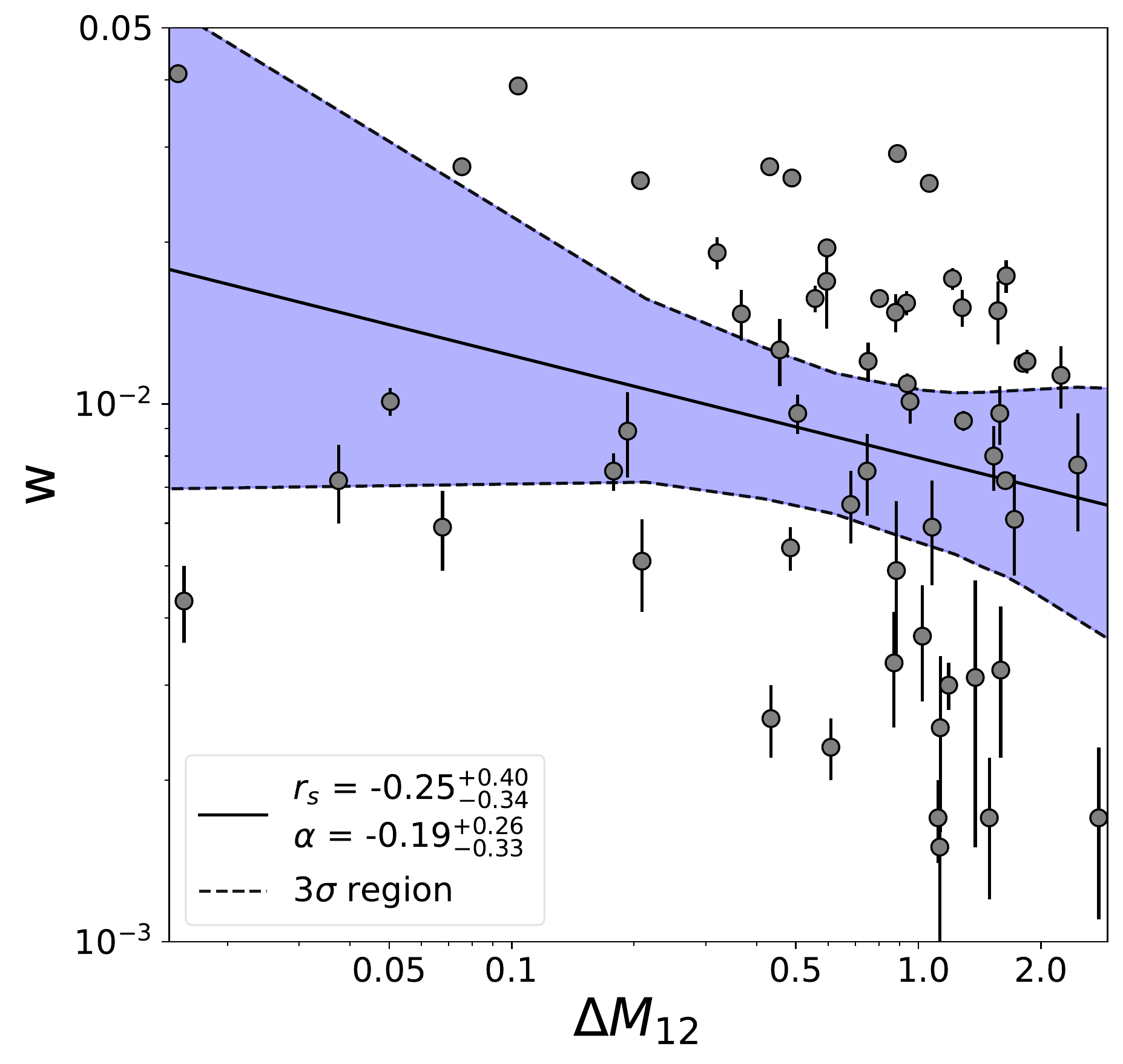}}
		\subfigure{\includegraphics[width=4.9cm]{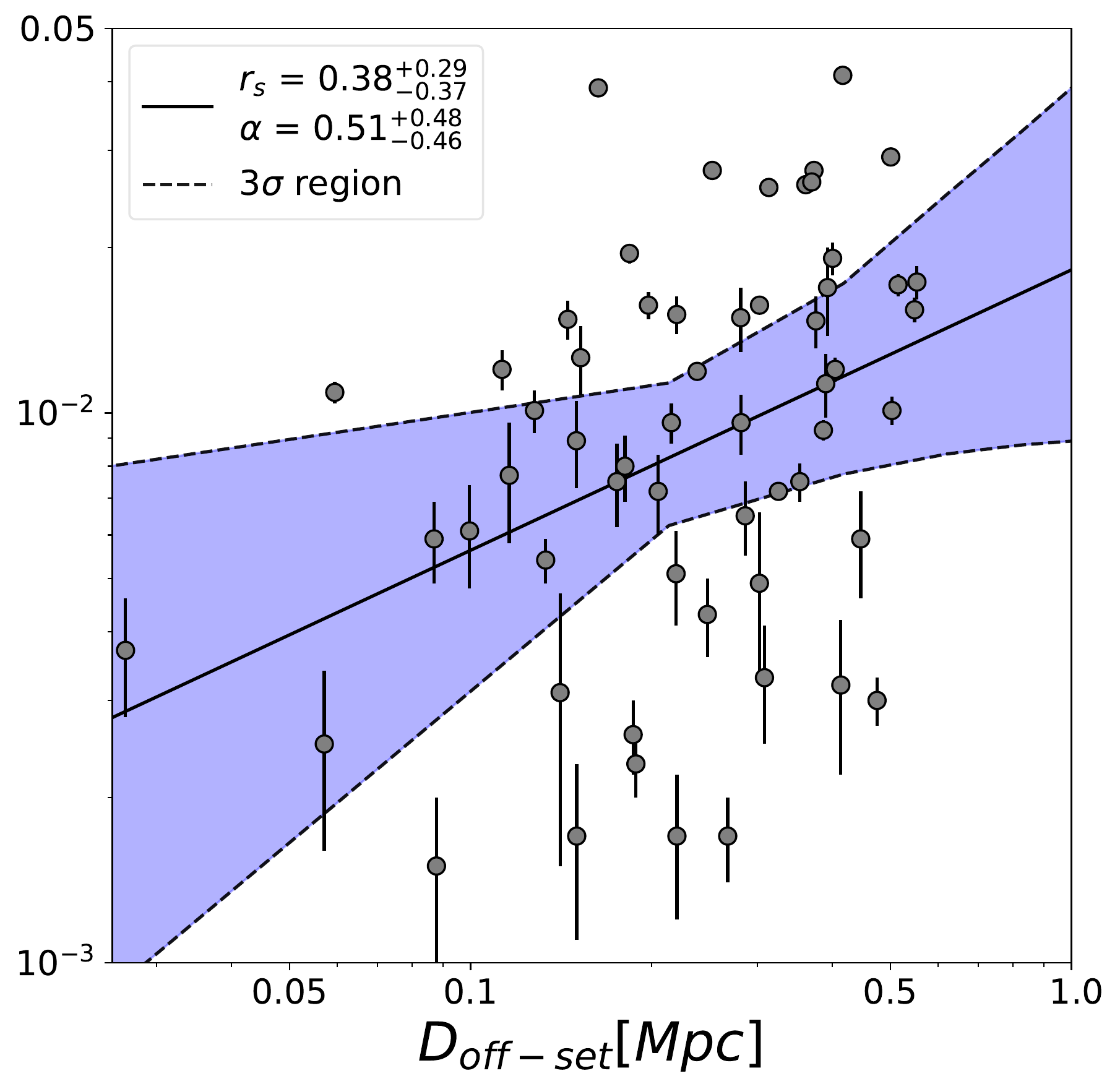}}
		\subfigure{\includegraphics[width=4.9cm]{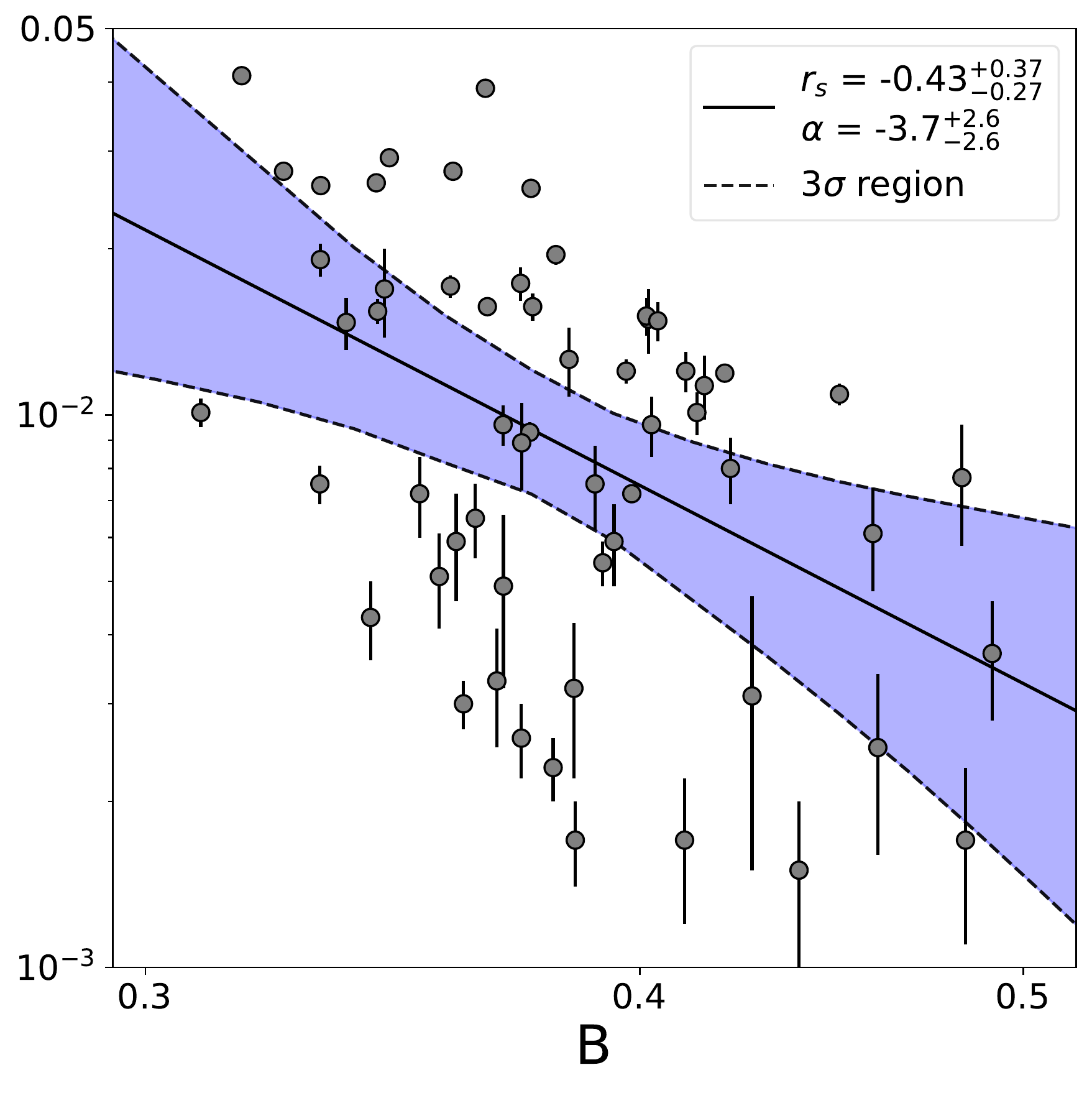}}
		
		\caption{X-ray probes versus photometric ($\Delta M_{12}$, $D_{\mathrm{off-set}}$, $B$) relaxation indicators. The solid line is the best-fit power-law relationship. The Spearman correlation coefficient, $r_s$, and the best-fit power-law slope, are indicated in each panel.The shaded area is showing a $3 \sigma$ confidence intervals of the fits.}
		\label{fig:aphot}
	\end{figure*}

	In summary, we show that the bivariant correlation , which is estimated by a combination of photometric indicators,  is better than luminosity gap or de-centering with a factor of 1.3. 
	In addition, our analysis and results are independent of the dynamical age definition ($\alpha_{1,0}$ or $\alpha_{0.5,0}$) as reported in Tables \ref{tabel1}, \ref{tabel05}.
	
	\subsubsection{Correlation between $\alpha_{z,0}$ and spectroscopic probes}\label{sec:sim:spec}
	
	In Figure \ref{fig:age2}, we show the distribution of spectroscopic indicators as a function of  $\alpha_{z,0}$. Similar to Figure \ref{fig:age1}, the x-axis of all left hand side panels are $\alpha_{1,0}$  and  the right panels' x-axis are $\alpha_{0.5,0}$. The solid straight line is the best fit line with $3 \sigma$ confidence interval and the blue curves is the median of $A^2$ and $\bar{\Delta V}$.  All panels comprise both scattered data and also the first(0.25) and third(0.75) quartiles ($Q_1 \& Q_3$), shown in the shaded regions.
	
	 In the top panels of Figure \ref{fig:age2}, we show the distribution of $A^2$ versus $\alpha_{z,0}$ using our own mock catalog explained in section \ref{mock}. The Spearman's correlation coefficient for $A^2 - \alpha_{1,0}$ is $r_s=-0.12^{+0.10}_{-0.09}$ and it is more or less similar to the correlation with $\alpha_{0.5,0}$ with $r_s=-0.10^{+0.09}_{-0.09}$.
	We see from top left panel of Figure \ref{fig:age2} that values of  $A^2$ at $\alpha_{1,0}= 0.3$ and $\alpha_{1,0}=0.5$ are $0.53$ and $0.46$, respectively.
		 We consider those halos with $A^2 > 0.53$ as unrelaxed/non-virialized and those with $A^2 < 0.46$ as relaxed/virialized systems. Based on our calculations, $39\%$ of virialized halos are early-formed and  $15\% $ are late-formed. Also, $33\%$ of non-virialized halos are early-formed and  $23\% $ of them are late-formed.
		 Moreover, although our results suggest that there is an anti-correlation between $\alpha_{z,0}$ and $A^2$, this anti-correlation is so weak that it can not properly determine the relaxation status of galaxy systems
	, and using $A^2$ can not bring about promising results in determining whether halos are relaxed or unrelaxed in our simulation study.	
	
			\begin{table}
		\centering
		\begin{tabular}{l l l l l }
			\hline
			Proxy &   slope    & intercept    &  $r_s$        & p-value        \\
			\hline
			$\log(A^2)$    & $-0.08^{+0.02}_{-0.03}$ & $-0.41^{+0.05}_{-0.06}$    & $-0.12^{+0.09}_{-0.09}$  & $0.01$  \\        
			$\log(\bar{\Delta V})$  & $-0.50^{+0.13}_{-0.13}$   &$0.80^{+0.33}_{-0.33}$   & $-0.39^{+0.07}_{-0.07}$   & 0.0  \\
			$\log (D_{off-set})$    & $-0.12^{+0.08}_{-0.07}$      & $-0.17^{+0.04}_{-0.04}$      & $-0.35^{+0.09}_{-0.08}$   & 0.0     \\
			$\log(\mathrm{B} )$    & $0.95^{+0.5}_{-0.4}$      & $-0.07^{+0.05}_{-0.03}$      & $0.47^{+0.08}_{-0.07}$    & 0.0    \\    
			$\log(\Delta m_{12})$    & $0.13^{+0.05}_{-0.04}$      & $-0.39^{+0.01}_{-0.02}$      & $0.35^{+0.08}_{-0.08}$     & 0.0       \\
			\hline        
		\end{tabular}
		\caption{Correlations between different relaxation proxies and $\log(\alpha_{1,0})$ (Unlike Table \ref{tabel1}, here we consider $\log(\alpha_{1,0})$ as a dependent variable and various proxies as independent variables).}
		\label{tabel4}
	\end{table}

	In the bottom panel of Figure \ref{fig:age2} we show the distribution of velocity segregation as a function of $\alpha_{z,0}$  to see the correlation between these two quantities. We demonstrate that the velocity-segregation is linked to the $\alpha_{1,0}$ with $ r_s=-0.43^{+0.08}_{-0.07}$ and $\alpha=-0.41^{+0.08}_{-0.09} $. The Pearson correlation coefficient for  $\alpha_{0.5,0}$ turn out to be $ r_s=-0.42^{+0.08}_{-0.08}$ in a similar way. In the bottom panel of Figure \ref{fig:age2} we reveal that the velocity-segregation of the dynamically relaxed groups is smaller than the velocity-segregation of unrelaxed groups confirming the observational study of \cite{Raouf2019b}.
	
	Finally, like the former arguments, we measure the range of growth histories that correspond to the velocity-segregation here. From bottom left panel, the values of  $\bar{\Delta V}$ at $\alpha_{1,0}= 0.3$ and $\alpha_{1,0}=0.5$ are  $\bar{\Delta V} = 253 kms^{-1}$ and $\bar{\Delta V} = 208kms^{-1}$ respectively. Like before, we consider those halos with $\bar{\Delta V} > 253 kms^{-1}$ as unrelaxed and those with $\bar{\Delta V} < 208 kms^{-1}$ as relaxed. Consequently,  $47\%$ of relaxed halos turn out to be early-formed and  $10\% $ are late-formed. On the other hand, $19\%$ of unrelaxed halos tend to be early-formed and  $39\%$  late-formed. 
	 So, by using the velocity-segregation, a large number of early-formed and late-formed halos could be correctly labeled as relaxed and unrelaxed respectively.
	
	In summary, when it comes to considering simulated data, we see that $\bar{\Delta V}$ provides a strong correlation with $\alpha_{z,0}$ compared to $A^2$. Moreover, we don't detect any dramatic distinction between using $\alpha_{1,0}$ or $\alpha_{0.5,0}$. For a better comparison between various age indicators discussed in this section, and more details of fitting parameters and correlation coefficients, we refer the reader to Tables \ref{tabel1}, \ref{tabel05}. As can be seen from the right panels of Figures \ref{fig:age1}, \ref{fig:age2}, the majority of data is distributed in the region where $\alpha_{0.5,0}>0.6$; as a result, the slope of the best fit line might be affected by the minority halos located at $\alpha_{0.5,0}<0.3$. This is not the case for the left-hand side panels as the data is more uniformly distributed in these plots, which are based on $\alpha_{1,0}$.

	It is worth mentioning that in all the above subsections, our null hypothesis is that there is no correlation between two variables, $H_0:r_s = 0$,  and our alternative hypothesis is the other way around (i.e: $H_1:r_s \ne 0$). Since the p-value for all cases are less than 0.05, we reject the null hypothesis with a significance level of 0.05. Additionally, as we are investigating the strength of the correlation, which is quantified by means of $r_s$, it doesn't differ to consider the above relaxation probes as a function of age or the other way around. In other words, even if we consider age as a function of the relaxation probes, there will not be any difference in our results since $r_s$ doesn't depend on the definition of the dependent and independent variable; however, the slope and intercept clearly changes (see Table \ref{tabel4}). The reported coefficients in Table \ref{tabel4} can be used to predict the age of a halo by means of the observable quantities.

		\begin{table}
		\centering
		\begin{tabular}{l l l l l }
			\hline
			Proxy &   slope    & intercept    &  $r_s$        & p-value       \\
			\hline
			$\log(A^2)$    & $0.98^{+0.91}_{-0.85}$   & $-0.96^{+0.32}_{-0.30}$      & $0.43^{+0.31}_{-0.38}$     &   0.001 \\        
			$\log(\bar{\Delta V})$  & $0.80^{+1.1}_{-1.2}$   & $-3.3^{+2.9}_{-2.9}$      & $0.31^{+0.35}_{-0.42}$   &  0.02 \\
			$\log(D_{off-set})$    & $0.68^{+0.91}_{-0.59}$   & $-0.83^{+0.53}_{-0.39}$      & $0.38^{+0.27}_{-0.36}$  &  0.05 \\
			$\log (\mathrm{B})$    & $-5.2^{+4.2}_{-3.8}$   & $3.4^{+1.6}_{-1.8}$      & $-0.43^{+0.27}_{-0.38}$    &  0.001  \\    
			$\log(\Delta m_{12})$    & $-0.26^{+0.43}_{-0.33}$   & $-1.3^{+0.21}_{-0.22}$      & $-0.24^{+0.42}_{-0.33}$    &  0.01 \\
			
			\hline        
		\end{tabular}
		\caption{Correlation between $\log(A_{phot})$  and different relaxation proxies. $A^2$ and $\bar{\Delta V}$ are spectroscopic indicators, while $D_{\mathrm{off-set}}$, $\mathrm{B}$ and  $\Delta m_{12}$ are photometric indicators. $3\sigma$ confidence intervals are calculated by the bootstrap method.}
		\label{tabel2}
	\end{table}

	\subsection{Relation between X-ray and optical indicators}
	
	We continue our analysis by examining the relationship between the X-ray and the optical indicators. Moving on to the photometric branch, we calculate $\Delta M_{12}$, $D_{\mathrm{off-set}}$, and $\mathrm{B}$ within the half virial radius for galaxy groups in Yang catalog \citep{Yang2005,Yang2007}. Also, when it comes to considering the spectroscopic probes, we calculate $A^2$ and $\bar{\Delta V}$ for this galaxy systems. We also compute the correlation between these quantities and the X-ray proxies in a similar way used in the simulation data. 
	
	\subsubsection{ Correlation of the X-ray indicators with photometric ones} 
	
	In Figure \ref{fig:aphot} we show the distribution of the X-ray indicators , which are discussed in section \ref{sec:x-ray:indic}, versus three different photometric probes. The y-axis of all top plots is in terms of  the photon asymmetry, $A_{phot}$; however, the bottom panel's y-axis is based on   the centroid shift, $w$. The solid straight line is the best fit line with $3 \sigma$ confidence interval illustrated in the shaded region.  All plots comprise observed data points with their error bars. For the sake of a simple comparison with results of \cite{Roberts2018}, in our figures we follow a similar plotting style.
	
	In the top left panel of this figure we show the distribution of $A_{phot}$ with measurement errors versus $\Delta M_{12}$. As can be seen , there is an anti-correlation with the coefficient of $r_s = -0.24^{+0.42}_{-0.33}$. 
	 It means that relaxed galaxy systems, which have smaller photon asymmetry, have larger $\Delta M_{12}$. In other words, this trend shows that by increasing $\Delta M_{12}$, $A_{phot}$ decreases. 
	In the top middle panel we show the $A_{phot}$ versus $D_{\mathrm{off-set}}$, which demonstrate a correlation with the coefficient of $r_s = 0.38^{+0.27}_{-0.36}$.  It shows that for unrelaxed systems that $A_{phot}$ is large, the $D_{\mathrm{off-set}}$ is large.

	Moving on our second X-ray indicator, in the bottom left panel of this figure we show the distribution of $w$ with the measurement errors as a function of $\Delta M_{12}$ . As can be seen there is  anti-correlation with coefficient of $r_s =-0.25^{+0.40}_{-0.34}$. It shows that relaxed groups/clusters, which have small centroid shift, have larger $\Delta M_{12}$.
		In the bottom middle panel we show the same figure for $D_{\mathrm{off-set}}$, which demonstrate a correlation with $r_s = 0.38^{+0.29}_{-0.37}$. Therefore, we can conclude that the more a galaxy system be relaxed, the smaller $D_{\mathrm{off-set}}$ and $w$ should be. 
		
		Finally, in the top right panel we plot $A_{phot}$ versus $\mathrm{B}$ defined in \ref{sec-bivariant}, which as a photometric indicator shows the strongest correlation with $A_{phot}$ with $r_s = 0.43^{+0.27}_{-0.38}$. It shed some light on this fact that unrelaxed systems have larger $\mathrm{B}$.
		Also, in the bottom right panel, we show the relation between  $w$ and $\mathrm{B}$. We find that this photometric indicator shows the strongest correlation with $w$ with $r_s = 0.43^{+0.28}_{-0.35}$. Consequently, it can be inferred that unrelaxed systems have larger $\mathrm{B}$.

	In summary
		, an interesting aspect of this results is that the bivariate correlation, $\mathrm{B} = 0.04 \times \Delta m_{12} - 0.11 \times Log(D_{off-set}) + 0.28$, is a prediction rather than a relation. In other words, this photometric probe of halo relaxation shows a strong correlation with the X-ray probes with a factor of 1.8 compared to $\Delta M_{12}$ and with a factor of 1.13 compared to $D_{\mathrm{off-set}}$ , so offering an economical replacement to expensive and relatively rare X-ray and spectroscopic proxies.
	
	\begin{figure*}    
		\centering
		\subfigure{\includegraphics[width=5.5cm]{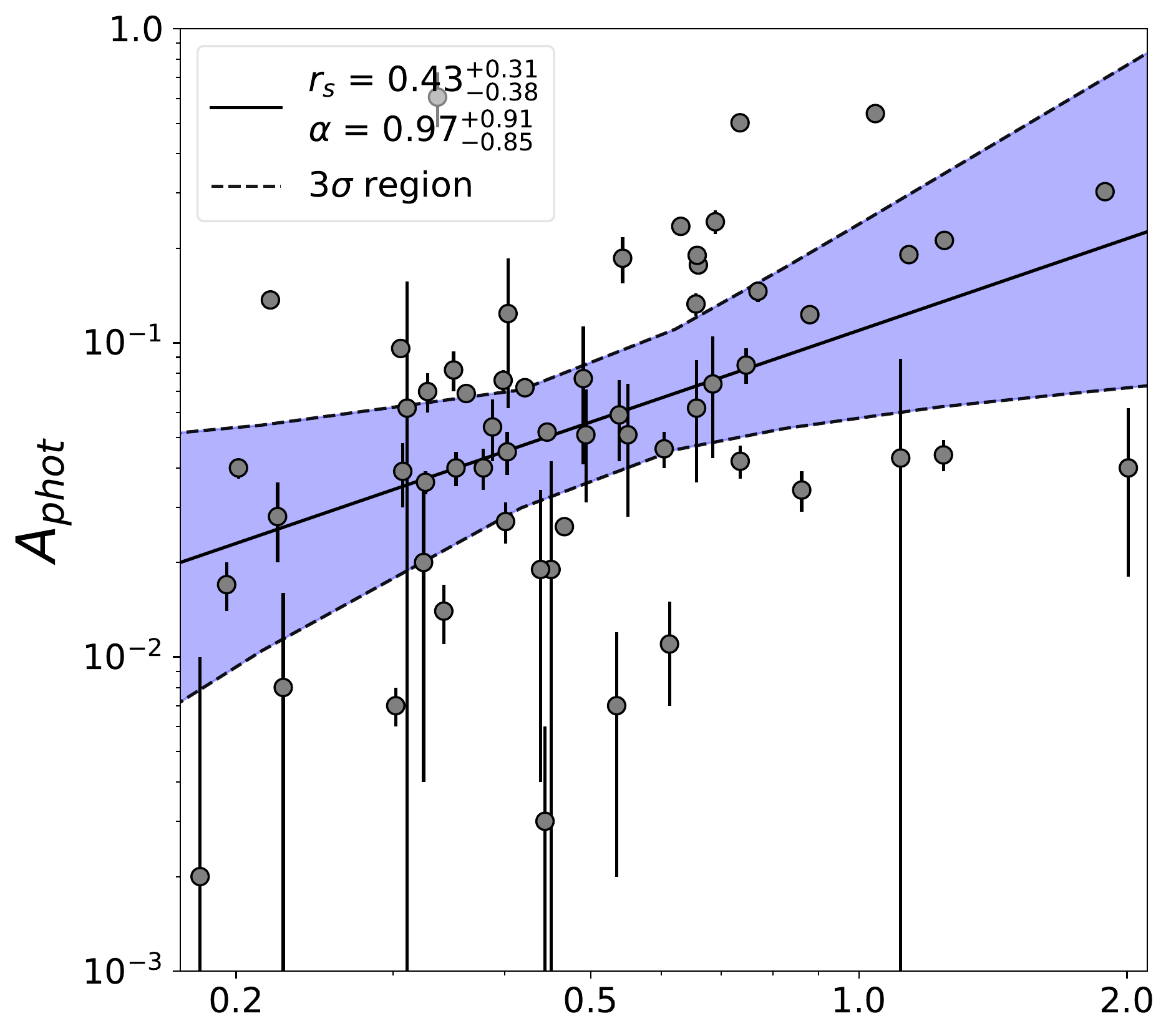}}
		\subfigure{\includegraphics[width=5.2cm]{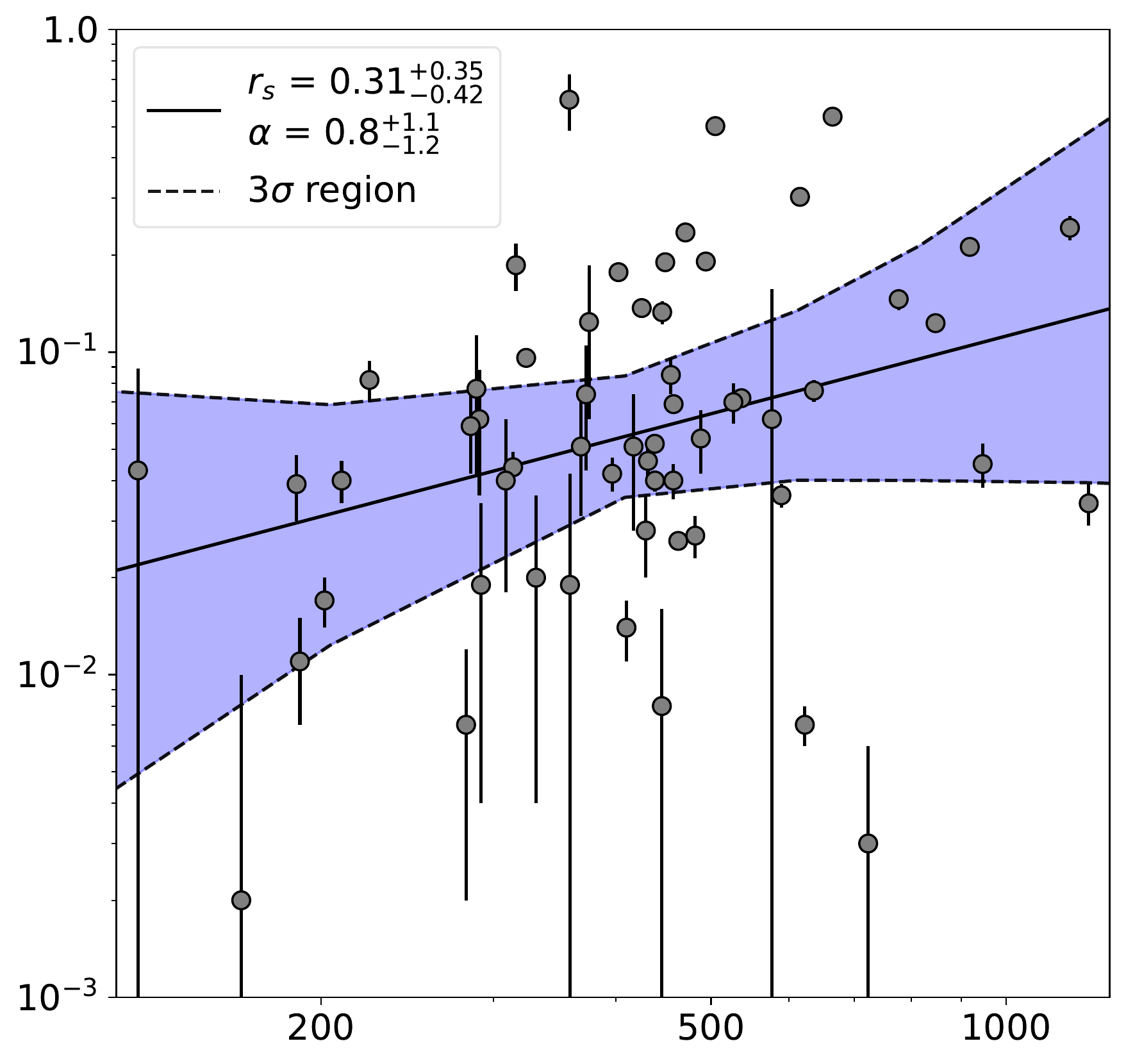}}
		\\[\baselineskip]
		\subfigure{\includegraphics[width=5.5cm]{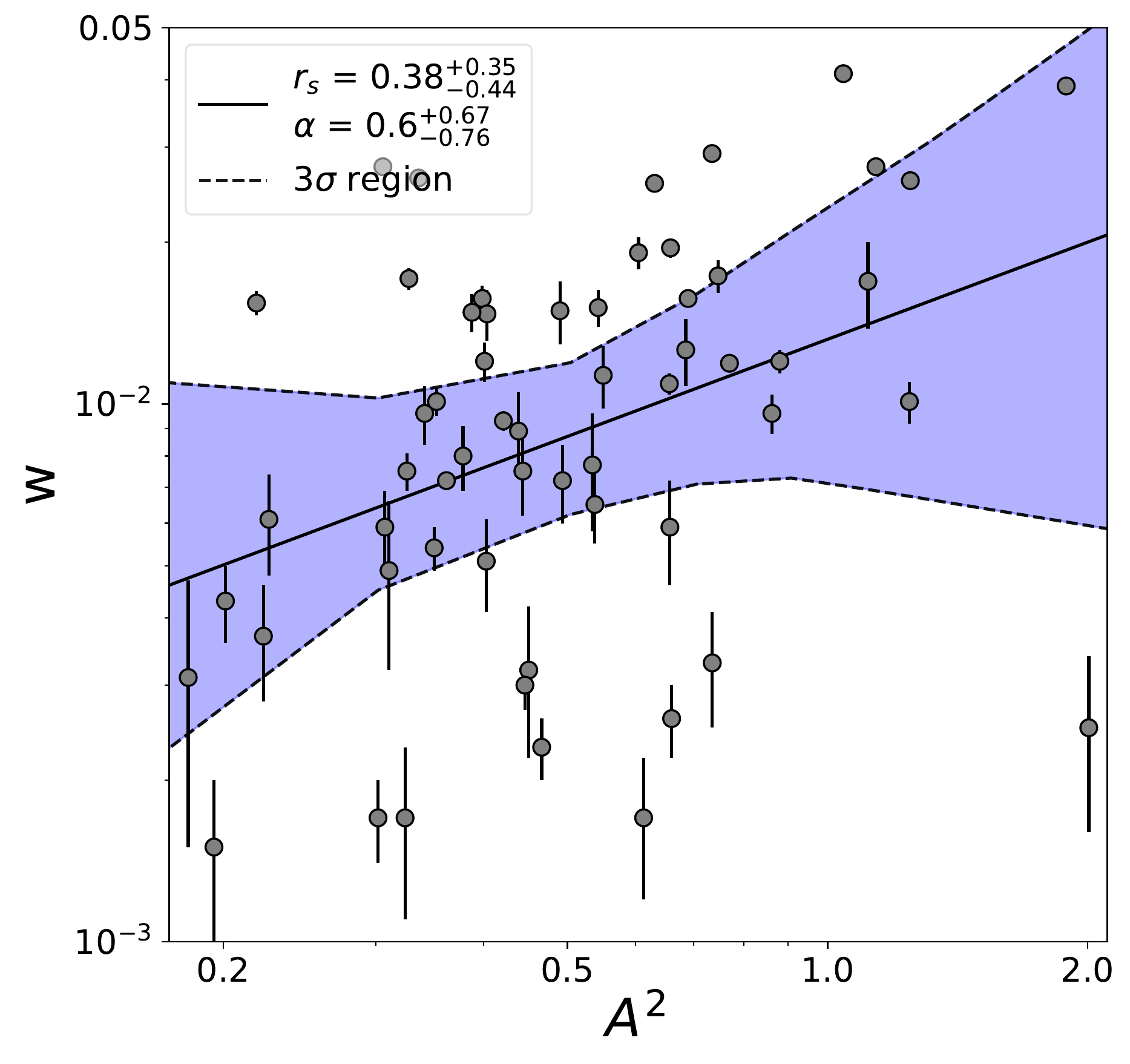}}
		\subfigure{\includegraphics[width=5.2cm]{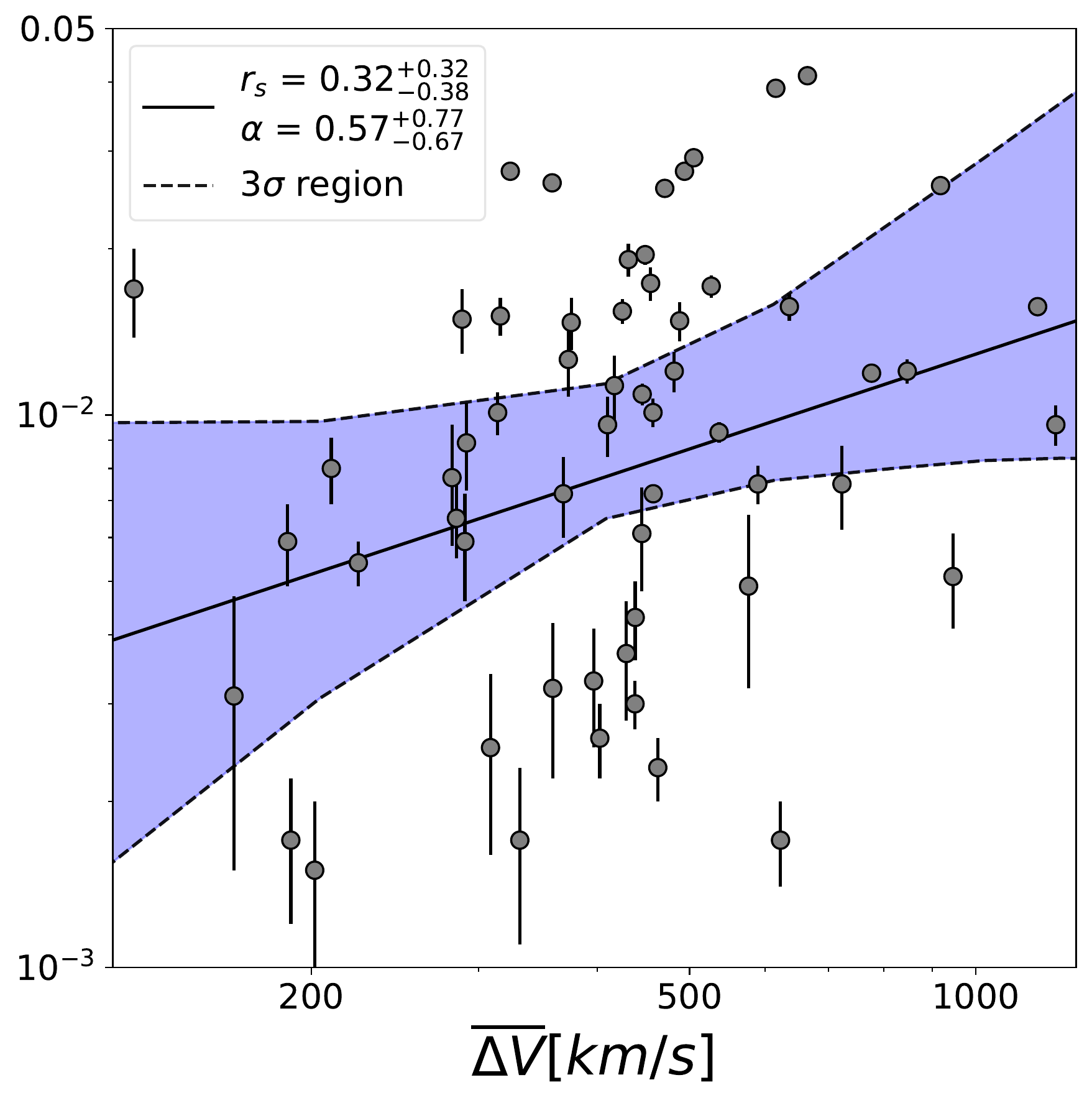}}
		\caption{ X-ray probes versus spectroscopic ($A^2$, $\Delta \bar{V}$) relaxation indicators. The solid line is the best-fit power-law relationship. The Spearman correlation coefficient, $r_s$, and the best-fit power-law slope, are indicated in each panel. The shaded area is showing a $3 \sigma$ confidence intervals of the fits.}
		\label{fig:w}
	\end{figure*}

	\subsubsection{Correlation of the X-ray indicators with spectroscopic ones} 
	
	In Figure \ref{fig:w} we illustrate the X-ray indicators versus spectroscopic probes. The y-axis of all top panels are in terms of $A_{phot}$; however, the bottom panels y-axis are based on $w$. The solid straight line is the best fit line with $3 \sigma$ confidence interval is illustrated in the shaded region.  All plots comprise observed data points with their error bars.
	
	In the top left panel of this figure, we plot $A_{phot}$ with the measurement errors versus $A^2$ . As can be seen there is a correlation with the coefficient of $r_s = 0.43^{+0.31}_{-0.38}$. It is consistent with our intuition that the velocity distribution of relaxed galaxy systems is supposed to be more normal than their unrelaxed counterparts.
	After that, in  the top-right panel we plot $A_{phot}$ versus $\bar{\Delta V}$, which demonstrates a correlation with coefficient of  $r_s = 0.31^{+0.35}_{-0.42}$. 
	
	In the bottom left panel of this figure we show distribution of $w$ with the measurement errors versus $A^2$ . As can be seen the correlation is positive with $r_s = 0.38^{+0.35}_{-0.44}$. In the bottom right panel we plot $w$ versus $\bar{\Delta V}$, which demonstrate a correlation with $r_s = 0.32^{+0.32}_{-0.38}$. As a result, apparently, relaxed systems show small velocity segregation while unrelaxed ones show larger segregation.

	In summary, while there is a considerable correlation between AD test and the X-ray indicators, there is no significant correlation between AD test and halo mass assembly at different redshifts (see Tables \ref{tabel1}, \ref{tabel05}). Furthermore, comparing the results of this section with what we find for simulated data in Sec. \ref{sec:sim:spec}, we can infer that the velocity-segregation has a stronger correlation with the mass assembly history at different redshifts than with X-ray indicators with a factor of $\sim$ 1.34.

	We would like to draw the reader's attention to this matter that the null hypothesis is rejected with a significance level of 0.05 in all the above correlations with X-ray indicators. It is because the p-values of all the above cases are either equal or less than  $0.05$. All fitting parameters along with correlation parameters are provided in Tables \ref{tabel2}, \ref{tabel3}

	\begin{table}
		\centering
		\begin{tabular}{l l l l l }
			\hline
			Proxy &   slope    & intercept    &  $r_s$        & p-value        \\
			\hline
			$\log(A^2)$    & $0.60^{+0.67}_{-0.76}$   & $-1.9^{+0.23}_{-0.27}$      & $0.38^{+0.35}_{-0.44}$    &  0.02  \\        
			$\log(\bar{\Delta V})$  & $0.53^{+0.77}_{-0.67}$   & $-3.6^{+1.7}_{-2.0}$      & $0.32^{+0.32}_{-0.38}$  &  0.05\\
			$\log (D_{off-set})$        & $0.50^{+0.48}_{-0.46}$   & $-1.7^{+0.32}_{-0.32}$      & $0.38^{+0.29}_{-0.37}$  & 0.004   \\
			$\log (\mathrm{B}) $    & $-3.7^{+2.6}_{-2.6}$   & $3.6^{+01.1}_{-1.1}$      & $-0.43^{+0.28}_{-0.35}$   &  0.003 \\    
			$\log(\Delta m_{12})$    & $-0.19^{+0.26}_{-0.33}$   & $-2.1^{+0.15}_{-0.15}$      & $-0.25^{+0.40}_{-0.34}$  &  0.05  \\
			
			\hline        
		\end{tabular}
		\caption{Correlation between $\log(w)$  and different relaxation proxies. $A^2$ and $\bar{\Delta V}$ are spectroscopic indicators, while $D_{\mathrm{off-set}}$, $\mathrm{B}$ and  $\Delta m_{12}$ are photometric indicators. $3\sigma$ confidence intervals are calculated by the bootstrap method.}
		\label{tabel3}
	\end{table}

	\section{SUMMARY and DISCUSSION}
	
	We use the Radio-SAGE galaxy formation model to study optical and spectroscopic relaxation proxies in galaxy groups and clusters. We propose a bivariant correlation between the luminosity gap and BGG offset, built entirely on photometric indicators, which demonstrates a strong correlation with the dynamical age of galaxy groups. To have fair comparison with observations, we have made a mock catalog for our simulations catalog. In the mock catalog we compute the line of sight velocity which is necessary for the calculation of velocity-segregation and AD statistics, $A^2$.
	For comparison, we use a sample of observational data to probe the reliability of the proposed bivariate relation in identifying the dynamically relaxed groups. Our findings can be summarized in the following:
	\begin{itemize}

		\item
		{
			We show that a combination of $D_{\mathrm{off-set}}$ and $\Delta m_{12}$ as a new photometric indicator,  bring about a strong Pearson's correlation coefficient with the dynamical age. After fitting the simulated data, we find the bivariant benchmark as a function of luminosity gap and de-centring ($\mathrm{B} = 0.04\times\Delta m_{12} -0.11 \times Log(D_{off-set}) + 0.28$).  This relation demonstrates an stronger correlation with $\alpha_{z,0}$ compared to just using luminosity gap or centroid shift individually.
		}

		\item
		{
		 From simulations, we show that the bivariant benchmark is more desirable than AD test. It is because the correlation between mass assembly and bivariant correlation is much more stronger than the correlation between AD test and mass assembly by a factor of $\sim 3.6$. We also show that by considering those halos with $\mathrm{B} < 0.41$ as unrelaxed and those with $\mathrm{B} > 0.45$ as relaxed,   $54\%$ of relaxed halos come true to be early-formed and  $10\% $ be late-formed. Also, $16\%$ of unrelaxed halos are early-formed and  $36\% $ are late-formed.
		}

		\item
		{
			From our fittings, we see  that using either $\alpha_{1,0}$ or $\alpha_{0.5,0}$  leads to slightly different slopes and intercepts. However, we observe no significant difference in the Pearson's correlation coefficients of these two cases. Therefore, we show that considering fast or slow growth doesn't have any impact on our analysis. Also, as the slope of the best fit lines associated with $\alpha_{0.5,0}$ are steeper than their counterpart, it can be inferred that they are more sensitive to the changes of the halo mass assembly.
		}

		\item 
		{ 
	    Using the observational data we show that a combination of $D_{\mathrm{off-set}}$ and $\Delta m_{12}$, which we define as bivariant correlation,  considerably improves the strength of the correlation coefficient between this indicator and the X-ray proxies like photon asymmetry and centroid shift.
	    Compared to using luminosity gap or de-centring individually, this new probe leads to stronger correlations.
    	}

		\item
		{
			From observations, through the velocity-segregation, there is a stronger correlation between bivariant benchmark and X-ray proxies. Also, when it comes to considering the correlation between AD statistics and X-ray probes, we notice that AD statistics leads to the same value as the one between bivariant benchmark and photon asymmetry. However, the correlation between AD statistics and centroid shift is weaker than the correlaton between the bivariant benchmark and centroid shift.
		}

	\end{itemize}
	
	  Through the analysis on simulations data, considering the AD statistics, the velocity-segregation has a stronger correlation with the halo mass assembly. In contrast, observational data shows that AD statistics has a stronger correlation with the X-ray indicators than the velocity-segregation. Furthermore, we show that while the AD statistics appears to be a good indicator for halo relaxation for the observed data, it is not a highly distinctive probe when applied to simulated data. One possible explanation is that there are no spatial X-ray information in the simulations to make a direct comparison with the observations. Therefore we rely on indirect halo relaxation indicators. The halo age in simulations is based on the dark matter, while in the observations is based on the baryonic component. Another factor which plays role is that all the halo virilization/relaxation indicators, photometric, spectroscopic and X-ray have drawbacks due to the projection effects. However, they are statistically useful methods for estimation of the dynamical age. Also, as the uncertainties in the data are quite big, simply fitting a linear line can not describe the real variation of the data; however, we add the linear regression to emphasis the general trend of the data.
	
	This study suggests that the photometric relaxation proxies are as well as, if not better than, their spectroscopic counterparts. We support this claim through investigating simulated data as well as observational data. Given the challenges of spectroscopic observations, the photometric data will works as good when it comes to quantifying the halo age. Consequently, it appears both economical and reliable to use the bivariant correlation $\mathrm{B}$ to attribute relaxation to galaxy systems.

	\section{Acknowledgment}
	We thank Ian Roberts for providing us with the X-ray data used in 
	\citep{Roberts2018}. His help has undoubtedly allowed this analysis to progress faster.

\end{document}